\documentclass[fleqn,usenatbib]{mnras}

\usepackage{newtxtext,newtxmath}

\usepackage[T1]{fontenc}
\usepackage{ae,aecompl}

\usepackage{lineno}

\usepackage{graphicx}	
\usepackage{amsmath}	
\usepackage{comment}

\newcommand{\be}{\begin{equation}} 
\newcommand{\ee}{\end{equation}}

\usepackage{soul}	
\usepackage{xcolor}

\title[Gamma rays from RX J1713--3946]{Particle acceleration at RX J1713--3946}

\title[Gamma rays and neutrinos from RX J1713--3946]
{Gamma rays and neutrinos from RX J1713--3946 in a lepto--hadronic scenario}

\author[Cristofari et al.]{
P. Cristofari,$^{1}$\thanks{E-mail: pierre.cristofari@obspm.fr}
V. Niro,$^{2}$
S. Gabici$^{2}$
\\
$^{1}$ LUTH, Observatoire de Paris, 5 place Jules Jansen, 92195 Meudon\\
$^{2}$ Universit{\'e} de Paris, CNRS, Astroparticule et Cosmologie,  F-75006 Paris, France \\
}

\date{Accepted XXX. Received YYY; in original form ZZZ}

\pubyear{2021}

\begin{document}
\label{firstpage}
\pagerange{\pageref{firstpage}--\pageref{lastpage}}

\maketitle

\begin{abstract}
The gamma--ray emission of RX J1713--3946, despite being extensively studied in the GeV and TeV domain, remains poorly understood. This is mostly because in this range, two competing mechanisms can efficiently produce gamma rays:  the inverse Compton scattering of accelerated electrons, and interactions of accelerated protons with nuclei of the ISM. In addition to the acceleration of particles from the thermal pool, the reacceleration of pre--existing CRs is often overlooked, and shall in fact also been taken into account. Especially, because of the distance to the SNR ($\sim 1$ kpc), and the low density in which the shock is currently expanding ($\sim 10^{-2}$ cm$^{-3}$), the reacceleration of CR electrons pre--existing in the ISM, can account for a significant fraction of the observed gamma--ray emission, and contribute to the shaping of the spectrum in the GeV--TeV range. Remarkably, this emission of leptonic origin
is found to be close to the level of the gamma--ray signal in the TeV range, provided that the spectrum of pre-exisiting cosmic ray electrons is similar to the one observed in the local interstellar medium.
The overall  gamma-ray spectrum of RX J1713--3946 is naturally produced as the sum of a leptonic emission from reaccelerated CR electrons, and a subdominant hadronic emission from accelerated protons.
We also argue that neutrino observations with  next--generation detectors might lead to a detection even in the case of a lepto-hadronic origin of the gamma-ray emission. 
\end{abstract}

\begin{keywords}
Stars: : general -- Interstellar medium: Cosmic Rays -- gamma-rays: general.
\end{keywords}


\section{Introduction}
Supernova remnants (SNRs) are crucial targets of interest for the gamma--ray community. Indeed, the gamma--ray emission in the GeV and TeV range originating from numerous SNRs is clearly demonstrating  that efficient particle acceleration is taking place. The detection of so far at least 30 SNRs in the GeV range~\citep{Fermi15}, and 12 SNRs in the TeV range~\citep{HESSSNR} is also a strong support to the idea that SNRs produce the bulk of Galactic cosmic rays (CRs). This idea, which is supported by several strong arguments, such as e.g. the fact that SNRs can inject into the ISM particles accelerated through diffusive shock acceleration (DSA) with a spectral distribution somewhat compatible with CR measurements at the Earth, and that they can account for the CR energy density at the Earth~\citep[see e.g.,][]{drury2012,blasi2013}, is however facing several obstacles~\citep[see e.g.,][]{tatischeff2018,gabici2019}. Let us for instance mention two major issues: 1) the fact that all detected SNRs seem to not be able to accelerate PeV particles, which is required for the sources of Galactic CRs; 2) our inability to clearly understand the mechanisms at stake in the production of gamma rays in the GeV and TeV range. Indeed, in these energy ranges, two mechanisms can efficiently produce gamma rays: an hadronic mechanism --the production of neutral pions in the interaction of accelerated protons with nuclei of the interstellar medium (ISM) subsequently decaying into gamma rays; and a leptonic mechanism-- the inverse Compton scattering of accelerated electrons with soft photons (CMB, IR, or optical). 

RX J1713--3946 is one of the best studied SNRs in the gamma--ray domain, and perfectly illustrates the difficulties to disentangle between the two possible mechanisms. In the literature, extensive discussions between the importance of the leptonic, hadronic mechanisms, or mixture of the two have been proposed. Although different scenarios have successfully managed to account for the observed gamma ray emissions, so far, a consensual and definitive interpretation of the high--energy emission is still missing~\citep{berezhko2008,morlino2009,fang2009,yamazaki2009,casanova2010,zirakashvili2010,ellison2012,dermer2013,yang2013,kuznetsova2019,tsuji2019,zhang2019,fukui2021}. 

It is commonly believed that a "peaked" shape of the gamma--ray spectrum tends to favor a leptonic mechanism~\citep{abdo2011,HESSRXJ}, which naturally produces a somewhat compatible shape.
However, in the case of RX J1713--3946 a leptonic gamma-ray spectrum would be too narrow if only one population of electrons accelerated at the shock is considered~\citep{finke2012}. 
Moreover, it has been pointed out that a hadronic origin can produce a good fit to the gamma--ray data, for a SNR shock expanding in a clumpy medium~\citep{fukui2012,gabici2014,celli2019}. 
 
Recently, several works have indicated that in addition to the acceleration of particles from the thermal bath at a SNR shock, the \textit{reacceleration} of already energized particles (i.e. CRs in the ISM) could also lead to a substantial contribution to the total gamma--ray emission. Especially, for SNRs expanding in a low density ISM (n$ \sim 10^{-2}$ cm$^{-3}$), the ram pressure of the shock ($\propto n v_{\rm sh}^2$) converted into CRs and the target density can be sufficiently low to provide a situation where the gamma rays from inverse Comption scattering of $\textit{re--accelerated}$ CR electrons becomes a significant contribution to the total gamma--ray emission~\citep{cristofari2019}.  Remarkably, such gamma--ray emission relies on only one assumption: the fact that the strong shock expands in a medium with a distribution of CRs that is the one measured at the Earth. It does not depend on the density in which the SNR shock expands, the velocity of shock, or an injection efficiency. The gamma--ray emission from reaccelerated electrons can therefore be seen as lower limit on the leptonic emission, that exists regardless of any property of the shock. 

In this context, we illustrate that the gamma--ray emission of RX J1713--3946 can naturally be accounted for taking into account the reaccelerated electrons, and an injection of protons from the thermal pool, straightforwardly producing the observed broad bump in the overall GeV -- TeV range. We additionally compute the number of neutrinos expected in the $\gtrsim 1$ TeV range in such mixed lepto--hadronic scenario, and illustrate that within 10-20 years of observations with next--generation instruments going beyond the km$^3$ volume \citep[e.g.][]{km3net2016} a detectable neutrino signal can still potentially be expected even in lepto-hadronic scenarios. 

The mixed lepto-hadronic scenario presented in this paper is based on the assumption that the spectrum of CR electrons pre-existing upstream of the SNR shock is identical to the one measured in the local ISM.
However, as the spatial distribution of CR electrons in the Galaxy is not well constrained, and might be characterised by significant spatial variations at both very large \citep{atoyan1995} and very low \citep{phan2018} particle energies, we conclude that neutrino observations of  RX J1713--3946 are mandatory in order to distinguish between different scenarios for the origin of its gamma-ray emission.

\section{Production of high--energy protons and electrons}
\subsection{Acceleration of particles from the thermal pool}
The acceleration of particles around the strong non--relativistic SNR shock waves (of compression factor $r=4$) expanding after the explosion of the parent supernova (SN) is assumed to be due to DSA, and is described with the usual assumption that a fraction  $\xi$ of the ram pressure $\rho v_{\rm sh}^{2}$ is converted into CRs, where $v_{\rm sh}$ is the shock speed. The CR proton spectrum at the shock is 
 $f^{\rm p}(p,t)=A(t) \left( \frac{p}{m_{\rm p}c}\right)^{-\alpha}$, with $\alpha=3r/(r-1)$ and the normalization $A$  reads $A(t)=(3/4\pi)\xi \rho v_{\rm sh}(t)^{2}/(m_{\rm p}^{4}c^{5}I(\alpha))$, with $I(\alpha)=\int_{p_{\rm min}/m_{\rm p}c}^{p_{\rm max}(t)/m_{\rm p}c}\text{d}x ~x^{4-\alpha}/(1+x^{2})^{1/2}$. This spectrum is exponentially suppressed at $p_{\rm max}(t)$. 
 
The spectrum of electrons accelerated from the thermal pool can subsequently be expressed as in~\citep{morlino2009,zirakashvili2007}:
\begin{equation}
f^{\rm e} (p,t) = K_{\rm ep} f^{\rm p}(p,t) \left[1 + 0.523 \left(  \frac{p}{p^{\rm e}_{\rm max} (t)}  \right)^{9/4} \right]^2 \exp \left[  -  \left( \frac{p}{p^{\rm e}_{\rm max} (t)} \right)^2\right]
\end{equation}
where $ K_{\rm ep}$ is the electron--to--proton ratio, and $p^{\rm e}_{\rm max}$ is the maximum momentum of electrons. 

\subsection{Reacceleration of pre--existing  CRs}
Pionner works on DSA already stressed the potential importance of the reacceleration of pre--existing energized particles in the medium in which the SNR shocks expands~\citep{bell1978b}. In the case of infinite plane shock, the formalism presented in detail in~\citet{blasi2004,blasi2017} is helpful to understand that the pre--existing CRs can be seen, in a stationary problem, as a boundary condition upstream infinity of the shock for the solution of the transport equation describing particles accelerated at the shock. Such boundary condition leads to an additional term, that described the particles reaccelerated at the shock, due to the presence of seed particles far upstream of the shock $f_{\infty}$. In such formalism, the reacceleration of pre--existing particles obtained,  in the absence of non--linear effects, is: 
\be
f_{\rm reac}(p)= \alpha \int_{p_0}^p \frac{\text{d} p'}{p'} \left( \frac{p'}{p} \right)^\alpha f_{\infty}(p')
\label{eq:freac}
\ee
Remarkably, the expression of the spectrum of reaccelerated particles at the shock depends on very few ingredients: the shock compression factor throught the parameter $\alpha=3r/(r-1)$, 
the minimum momentum above which reacceleration occurs $p_0$, and of course, the presence of seed CRs $f_{\infty}$. 
In the case of a strong shock and in the test particle limit, $\alpha=3r/(r-1)$ is known and equal to 4. 

As discussed in~\citet{blasi2017}, $p_0$ is of little importance for the shape and normalization if the considered seeds are Galactic CRs, and provided that $p_0$ is sufficiently low. Indeed, for momenta below $\sim 1$ GeV/c, the electron and proton spectra are harder than $p^{-4}$, therefore, the integral over momentum in Eq.~\eqref{eq:freac} is dominated by momentum $\sim mc$ at low momenta, and by larger momenta above $mc$. In the following we will assume that $p_0=p_{\rm inj}= 10^{-2}$ mc. 
In this work, our main assumption will therefore be that  the seed electrons and protons in which RX J1713--3946 expands are the Galactic CRs. We will used parametrized descriptions of the unmodulated CR spectra provided for Galactic protons and electrons in~\citet{bisschoff2019}. These description are in good agreement with data collected by Voyager~\citep{cummings2016} or PAMELA~\citep{adriani2011,adriani2011b}, and we introduce a hardening in the proton and electron spectra $\propto p^{0.1}$ and $\propto p^{0.2}$, at 300 GeV for protons and 100 GeV for electrons to fit the  AMS--02 data~\citep{aguilar2015,aguilar2019}, as in~\citet{cristofari2019}. We assume that the presence of the stellar wind bubble does not significantly affect the ambient CR spectrum. This assumption is reasonable since the liftetime of the wind bubble is $\sim$ Myr, for a typical size of few tens of parsec, and, at $\sim$ GeV the typical diffusion length is $\sim$ a few hundreds of parsec, therefore giving enough time for CRs to populate the cavity.

 \subsection{Particles accelerated at RX J1713--3946}
 The particles (protons and electrons) accelerated, and reaccelerated at the shock can either be advected downstream of the shock, or manage to escape upstream of the shock into the ISM. In order to escape the SNR accelerator, particles must be sufficiently energized, so that the flux of escaping particles is often described as a delta function peaked around $p_{\rm max}(t)$. In this work, we do not consider this flux of escaping particles, as the gamma--ray emission of the SNR is largely dominated by the flux of the particle advected downstream of the shock and that we consider to be trapped  downstream of the expanding shock. From the beginning of the free expansion phase $t_0$ to a time $T$ , the total number of particles trapped downstream of the shock reads: 
 \be
 N(p)= \int_{t_0}^T \text{d}t \frac{4 \pi}{r} r_{\rm sh}^2(t) v_{\rm sh}(t) \left[ f(p) +f_{\rm reac}(p)\right]
 \ee
In the case of RX J1713--3946, we adopt $t_{0}=1$ yr and $T=1623$ yr. 

\subsection{Energy losses}
The trapped particles suffer adiabatic losses and synchrotron losses (only relevant for electrons). These losses can be taken into account by writing the conservation of the total number of particles inside the SNR. Following a particle of momentum $p'$ accelerated at a time $t$ and whose momentum is $p$ at $T$, the total number of particles reads: 
\be
N_{\rm loss}(p)= \int_{t_0}^T \text{d}t \frac{4 \pi}{r} r_{\rm sh}(t)^2 v_{\rm sh}(t) \left( \frac{p'}{p}\right)^2 \left[ f(p) + f_{\rm reac}(p) \right] \frac{\text{d}p'}{\text{d}p}
\ee
where the changing of momentum is given by: 
\be
\frac{\text{d}p}{\text{d}t}= - \frac{p}{{\cal L}} \frac{\text{d}{\cal L}}{\text{d}t} + \frac{4}{3} \sigma_{\rm T} c \left( \frac{p}{m_{\rm e}c} \right)^2 \frac{B_{\rm down}^2}{8 \pi}
\ee
with $\sigma_{\rm T}$ is the Thompson cross--section and ${\cal L}=[\rho_{\rm down}(t)/\rho_{\rm down}(t')]^{1/3}$ accounts for the adiabatic expansion between a time $t$ and $t'$~\citep[see e.g.,][and reference therein for more details]{cristofari2020,cristofari2021}.

 \subsection{Maximum momentum of accelerated particles}

The question of the maximum momentum of the accelerated protons and electrons  is essential in this problem, as the corresponding cut--off can shape the gamma--ray spectrum. Most updated results indicate that the maximum momentum of particles is dictated by the growth of magnetic instabilities, and instabilities growing with the fastest growth rate shall dominate the process. The fastest growing modes are expected to be non--resonant hybrid modes as discussed in~\citep{bell2004}. 
 In this case, the maximum momentum of particles is then set by the saturation of the mechanism, typically reached when the growth corresponds to a few (${\cal N}$) e-folds. If $\gamma_{\rm max}$ is the growth rate at the wavenumber where the growth rate is the highest, the saturation condition is: $\int_0^{t} \text{d}t' \gamma_{\rm max}(t')\sim {\cal N}$. The number of e-folds met at saturation ${\cal N}$ is yet poorly constrained. Indeed, typical values are inferred from numerical studies ${\cal N}\approx 5 $~\citep{bell2013}, but arguments in favors of values in the range $\sim 3 - 9$ can be made~\citep[see e.g., for detailed discussions][]{schure2014}. The different values of ${\cal N}$ could help explain particle acceleration up to the $\sim$ PeV range, or help understand the values of magnetic field energy density at SNRs derived from observations~\citep{volk2005}. 
 
 Assuming a typical value ${\cal N}=5$, one gets~\citep{bell2013}: 
\begin{equation}
\label{eq:pmax}
p_{\rm max}(t) \approx \frac{3 r_{\rm sh}(t)}{10} \frac{\xi e \sqrt{4 \pi \rho(t)} }{\Lambda} \left(\frac{v_{\rm sh}(t)}{c}\right)^{2},
\end{equation}
where $\Lambda=\ln\left(\frac{p_{\rm max}(t)}{mc}\right)$. 
The corresponding amplified magnetic field reads: 
\be
\delta B \approx 2 \sqrt{3 \pi \frac{v_{\rm sh}}{c}\frac{\xi\rho v_{\rm sh}^{2}}{\Lambda}}
\label{eq:satura}
\ee

When the SNR shock enter the low density bubble, the amplification of  the magnetic field through non--resonant growth of instabilities becomes inefficient. We consider that the magnetic field in the bubble is $\approx 5 \mu$G. The corresponding maximum energy of protons is then estimated using the Hillas criterion, equating the Bohm--like diffusion coefficient to a fraction $\chi \approx 0.05-0.1$ of the shock radius. 

Unlike protons, the maximum energy of electrons is affected by energy losses. These can be taken into account equating the time $\tau_{\rm acc}$ to the minimum of the loss time $\tau_{\rm loss}$ and the age of the system. The acceleration time is estimated as~\citep{drury1983}: 
\be
\tau_{\rm acc}= \frac{3}{v_1- v_2} \int_0^p \frac{\text{d}p'}{p'} \left(  \frac{D_1(p')}{v_1} + \frac{D_2(p')}{v_2}\right)
\ee
where indices 1 (2) refer to the region upstream (downstream) of the shock, $v$ is the fluid velocity, so that $v_1=v_{\rm sh}$ and $v_2=v_{\rm sh}/r$, and $D$ the diffusion coefficient assumed to be Bohm--like. As illustrated in~\citet{cristofari2019} assuming a different energy--dependence for the diffusion coefficient could result in a reduced maximum energy for the electrons. 
Finally, we assume that the maximum energy reached by accelerated and reaccelerated particles is the same.

\subsection{Dynamics of the SNR shock}
We assume that RX J1713--3946 is the remnant of a massive star that led to the explosion of a core collapse SN. Therefore the environment in which the SNR shock expands is structured by the history of the parent massive star: in its main sequence, the stellar winds inflates a low density bubble in pressure balance with the ISM. When entering the red super giant (RSG) stage, the low velocity dense wind forms, of typical velocity $u_{\rm w}=10^6$ cm/s, mass loss rate $\dot{M}=10^{-5}$ M$_{\odot}$/yr, and density $n_{\rm w}=\dot{M}/(4 \pi m u_{\rm w} r^2)$~\citep{weaver1977}. When the SN explosion occurs, the SNR shocks thus successively expands through the dense RSG wind, the low density bubble and finally reaches the unperturbed ISM. The transition between the dense RSG wind and low density cavity is set by the pressure equilibrium, and occurs at $r_1=\sqrt{\dot{M}u_{\rm w}/4 \pi k n_{\rm b} T_{\rm b}}$, where $k$ is the Boltzmann constant, the density and temperature of the hot low density cavity is $n_{\rm b}=2 \; 10^{-2}$~cm$^{-3}$, $T_{\rm b}=10^6$~K in the case of RX J1713--3946. 
In such environment, the dynamical evolution of the SNR shock can be described under the thin--shell approximation~\citep{BG1995,ptuskin2005}.

\section{High--energy observable messengers}

\begin{figure}
\includegraphics[width=.5\textwidth]{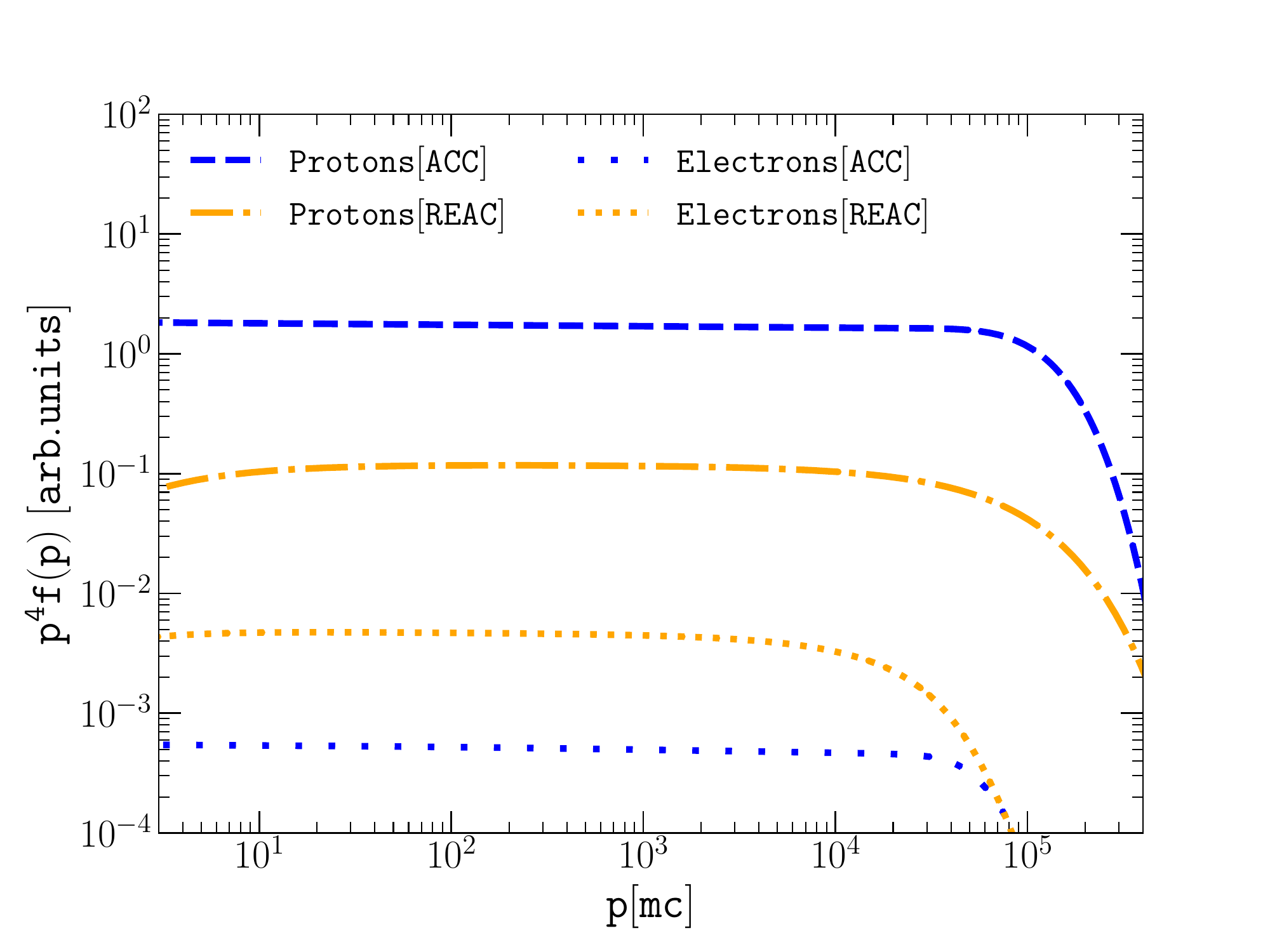}
\caption{Distribution of particles accelerated at  RX J1713--3946. Accelerated protons and electrons are shown in blue dashed and blue loosely dotted. Reaccelerated protons and electrons are shown in orange dot--dashed and orange dotted.}
\label{fig:particles}
\end{figure}

The distribution of protons and electrons,  accelerated and reaccelerated at RX J1713--3946 are shown in Fig.~\ref{fig:particles}. The proton content (accelerated and reaccelerated) dominates over the electron one (reaccelerated and electrons). However, because of the different mechanisms of production of gamma rays, we show that the gamma--ray signal from reaccelerated electrons (interacting with the photon fields) and from accelerated protons (interacting with the ISM) can be at around the same level.

\subsection{Radiations from non--thermal particles}

The non--thermal protons and electrons accelerated (and reaccelerated) can produce gamma rays through two mechanisms: pion production in proton--proton (pp) collisions, and inverse Compton scattering (ICS) of electrons on soft photons. The ICS contribution is estimated considering three Galactic photon fields: the Cosmic Microwave Backgorund, far--infrared dust emission and near--infrared stellar emission. These three components are assumed to be blackbodies of temperature 2.72K, 30K and 3000K and energy densities of 0.261, 0.5, and 1 eV/cm$^{3}$.

\begin{figure}
\includegraphics[width=.5\textwidth]{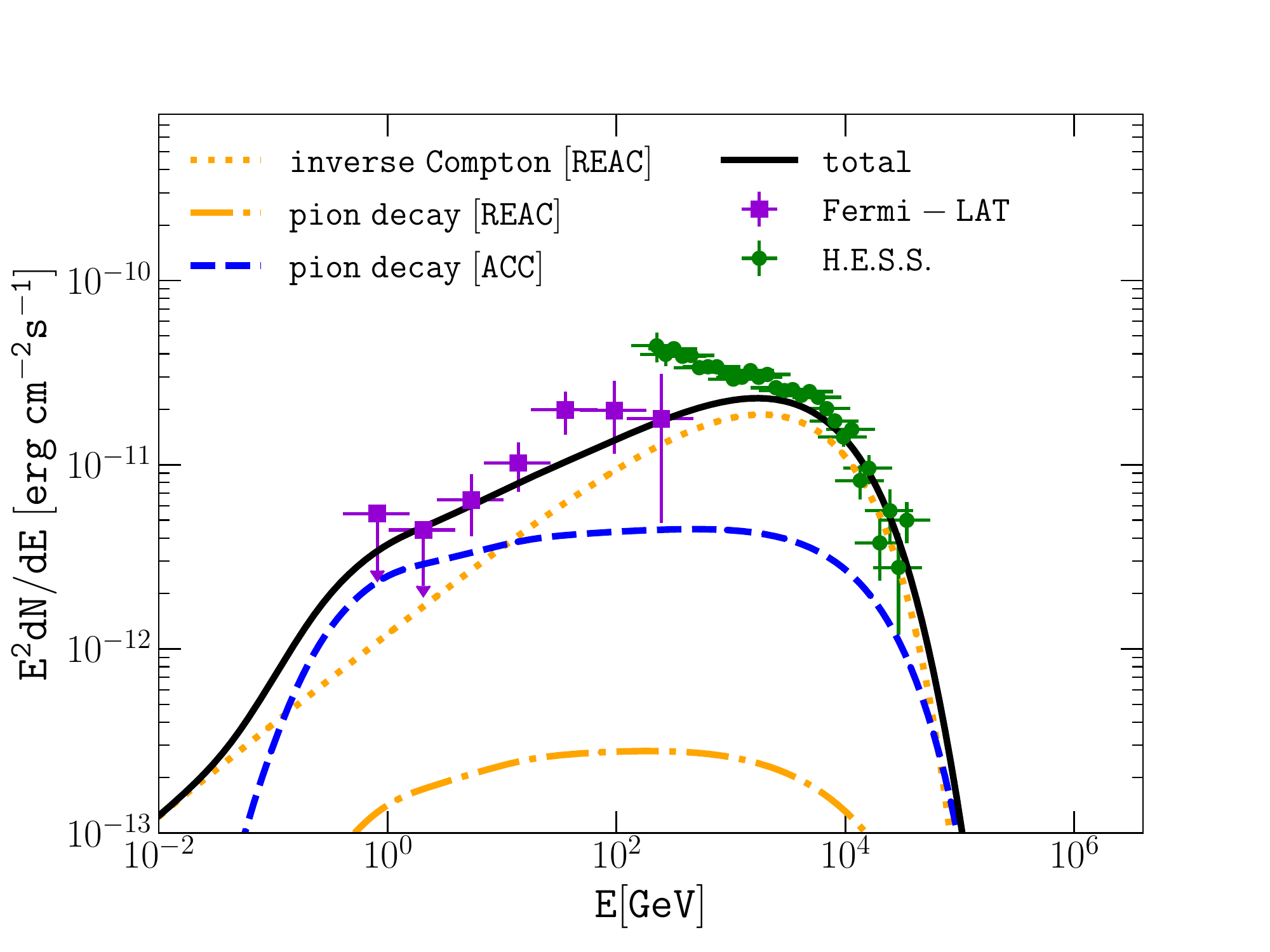}
\caption{Differential spectrum of RX J1713--3946 obtained with H.E.S.S.~\citep{HESSRXJ}  and Fermi--LAT~\citep{RXJFERMI} observations.  The dotted (yellow) and dot--dashed (yellow) lines correspond to the gamma rays from re--accelerated electrons and protons. The dashed blue line correspond to freshly accelerated protons.
The solid black line is the sum of gamma rays from freshly accelerated protons and re--accelerated electrons. }
\label{fig:spectrum}
\end{figure}

\begin{figure}
\includegraphics[width=.5\textwidth]{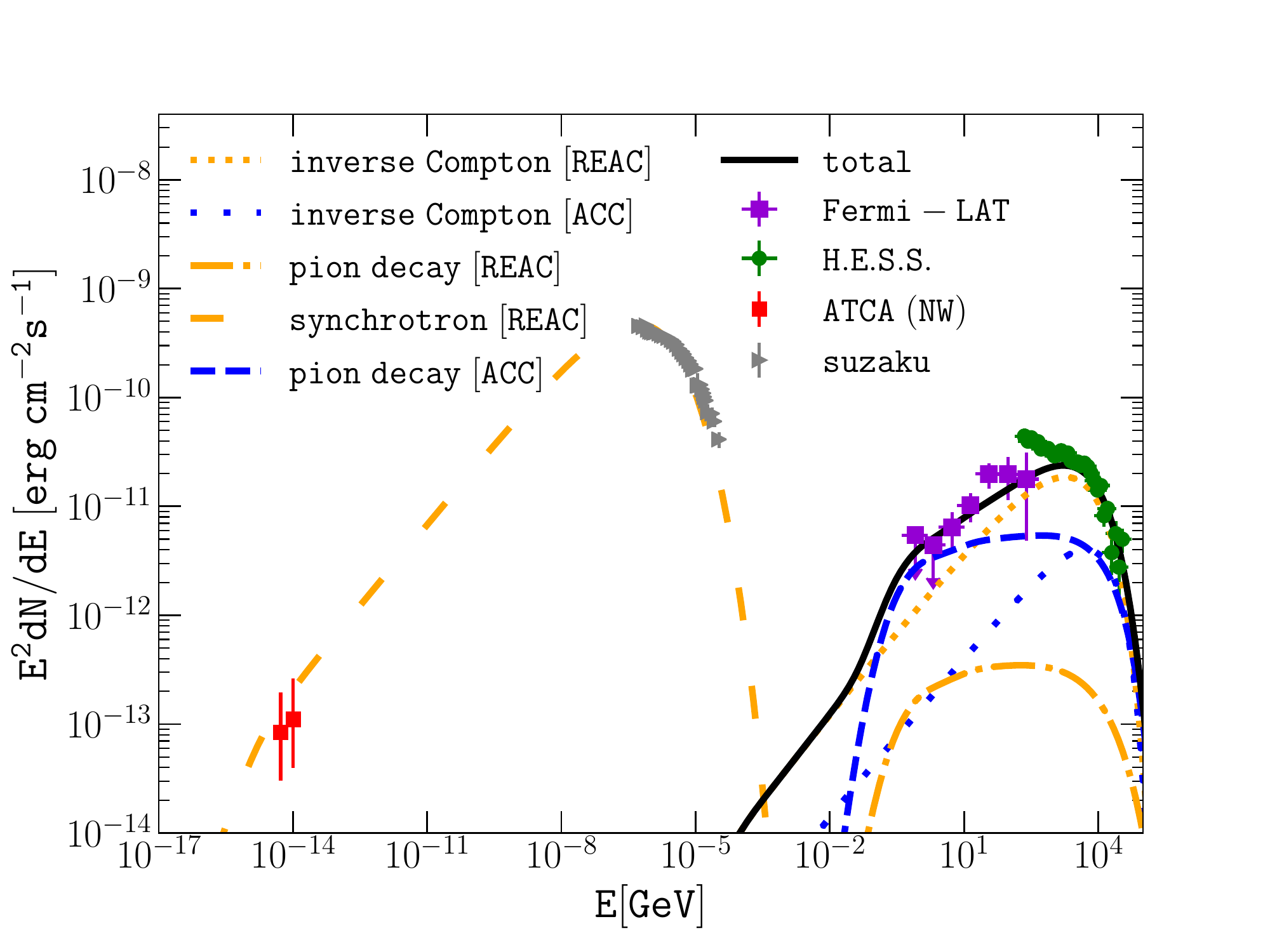}
\caption{Multi--wavelength differential spectrum of RX J1713--3946. Radio and X--ray observations were obtained from ACTA~\citep{ATCARXJ} and Suzaku~\citep{SUZAKURXJ}. Lines are as in Fig.~\ref{fig:spectrum}.}
\label{fig:large_spectrum}
\end{figure}

The corresponding gamma rays can be calculated as in~\citet{kelner2008} and~\citet{blumenthal1970}, for instance using the Naima package presented in~\citep{khangulyan2014}. The obtained gamma--ray spectrum is shown in Fig.~\ref{fig:spectrum}. The total gamma--ray differential flux is especially shown (black solid), and naturally exhibits a large bump, at the level of the Fermi--LAT and H.E.S.S. signals. 
The total emission in the $\sim$ TeV range is dominated by gamma--ray from reaccelerated electrons (dotted yellow line), while the pion decay emission from reaccelerated protons is subdominant (dot-dashed yellow line). On the other hand, the GeV part of the spectrum is dominated by the pion decay emission from freshly accelerated protons (assuming an acceleration efficiency of $\sim 10$\%, compatible with the idea that SNRs are the sources of Galactic CRs). 
The amount of electrons freshly accelerated at the SNR shock is usually accounted for by introducing an electron--to--proton ratio $K_{\rm ep}$ so that $f^{\rm e}(p)=K_{\rm ep} f^{\rm p}(p)$. $K_{\rm ep}$ is expected to be in the range $10^{-5}- 10^{-2}$, but it not well constrained~\citep{cristofari2013}. In order to explain the gamma--ray emission from freshly accelerated electrons, a typical value of $K_{\rm ep}\approx10^{-2}$ is needed. 
In this work, we adopt for illustrative purposes the value $K_{\rm ep}=10^{-4}$ (Fig.~\ref{fig:large_spectrum}), that would lead to a subdominant contribution from freshly accelerated electrons, although higher values of $K_{\rm ep}$ are possible, and would reinforce the signal from reaccelerated electrons.

The synchrotron emission from electrons trapped downstream of the SNR shock is computed in the average magnetic field inside the SNR: $<B_{\rm down}> = 1/V \int_{0}^{r_{\rm sh}(T)} \text{d}r_{\rm sh}4 \pi \; r_{\rm sh}^2 B_{\rm down}(r_{\rm sh}) \approx 16~ \mu $G, and is found to lead, in the X--ray and radio domains, to a signal at the level of the measured emission. The overall spectrum from the radio to the high--energy gamma--ray domain is shown in Fig.~\ref{fig:large_spectrum}.

\subsection{Neutrinos}
The source is located in the southern sky with a declination of~$-38.24^\circ$. 
For this reason, we investigate the possible detection of RX J1713--3946 with the KM3NeT instrument. In this section, we report the calculation for the neutrino events, using the KM3NeT effective area. 
Following the procedure outlined in~\citet{Niro:2019mzw}, we will use the effective area for the detector as reported in~\citet{Adrian-Martinez:2016fdl}. 
The event rate reads: 
\begin{equation}
N_{\rm ev}=\epsilon_v t\int_{E^{th}_\nu} \text{d}E_\nu \frac{\text{d}N_\nu(E_\nu)}{\text{d}E_\nu} A^{\rm eff}_{\nu}\,,
\end{equation}
where the parameter $\epsilon_v$= 0.7 is the visibility of the source. The background of atmospheric neutrinos~\citep{Volkova:1980sw,Honda:2011nf,Gondolo:1995fq} is then integrated over an opening angle equal to $\Omega = \pi \sigma^2_{\rm ext}$, with 
$\sigma_{\rm ext}= 0.65^\circ$~\citep[see e.g.,][]{Aharonian:2005qm,Abdalla:2016vgl}. The number of neutrino events from the source is calculated considering the expressions given in~\citet{Kappes:2006fg}\citep[see also][for a detailed description of the formulae]{Gonzalez-Garcia:2013iha,Halzen:2016seh}  and~\citet{Villante:2008qg} for an alternative derivation. Note that the formulae by~\citet{Kappes:2006fg} take into account neutrino oscillations, assuming full neutrino mixing. We found that, assuming that the gamma rays detected by Fermi--LAT and H.E.S.S. are entirely due to hadronic interactions, about $\sim$1.7 events are expected in one year of observation with KM3NeT and $\sim$0.7 background events considering an energy threshold of 1~TeV. Our results are in agreement with the one found in \citet{Costantini:2004ap} and \citet{Vissani:2011ea}, if the results reported in these references are rescaled by the visibility \citep[see also][for discussions on this topic]{Kappes:2006fg}. 

Instead, considering the mixed scenario of leptonic and hadronic emission described above (and illustrated in Fig.~\ref{fig:spectrum}), about 0.3 events are expected in one year running of the KM3NeT detector for particle energies exceeding 1~TeV.
In ten years of observation, this would correspond to about 3 signal neutrinos versus 7 backgrounds events above 1 TeV. 

\begin{figure}
\includegraphics[width=.45\textwidth]{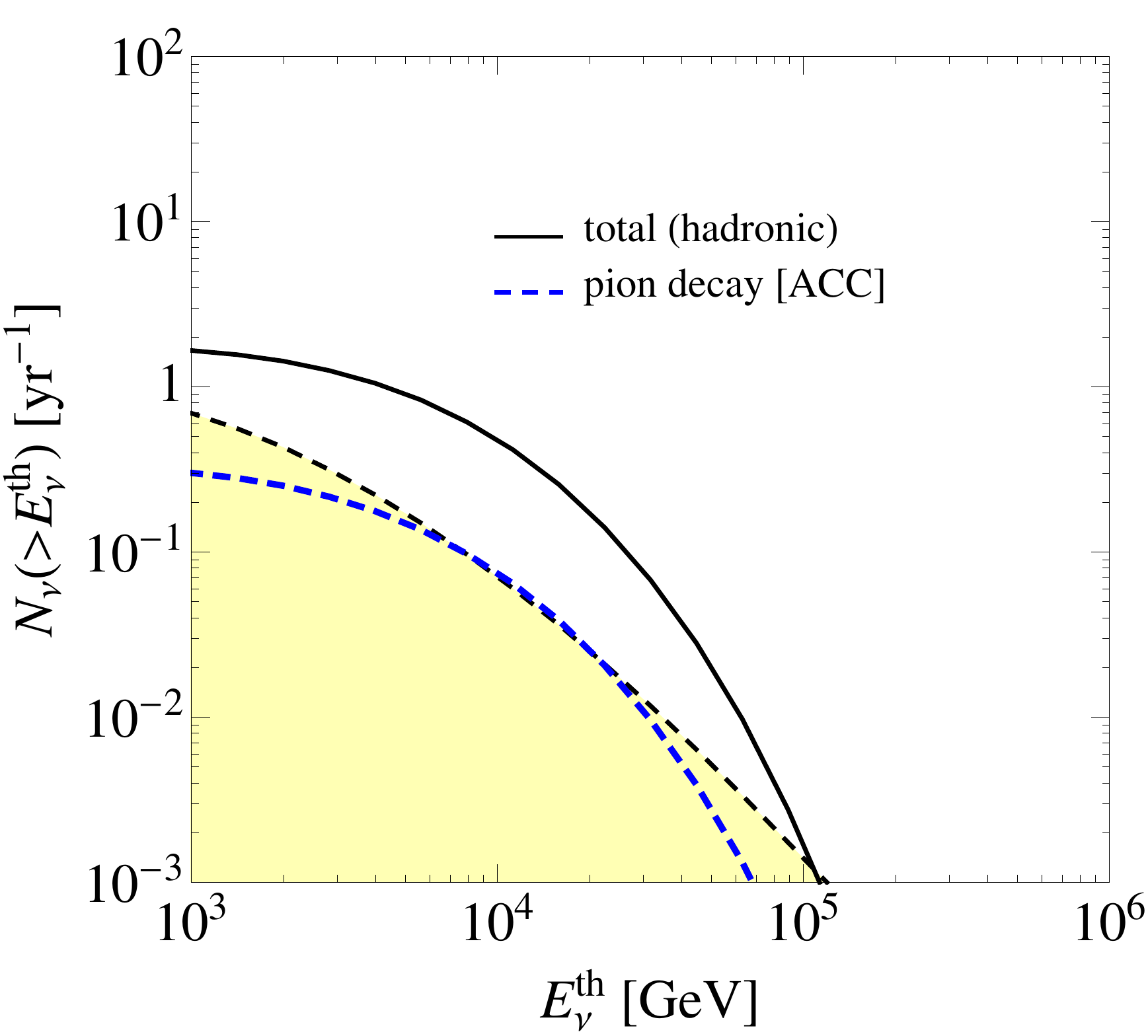}
\caption{Number of neutrino events at the KM3NeT detector from the RX~J1713.7--3946 source compared to the number of events expected by the atmospheric background (yellow shaded area). The black lines refer to the fully hadronic scenario, while blue lines to the lepto-hadronic case described in this paper (dashed). 
}
\label{fig:events}
\end{figure}

We report in Fig.~\ref{fig:events} the number of events as a function of the neutrino energy threshold for the two scenarios examined above: the fully hadronic (solid line) and the lepto-hadronic (dashed line). 
The expected number of background events is also shown as a shaded region. 
We see from Fig.~\ref{fig:events} that in the lepto--hadronic scenario, the number of signal events is of the same order of the background ones for particle energies above $\sim 10$ TeV (of the order of 0.1 neutrinos per year).

\begin{figure}
\includegraphics[width=.45\textwidth]{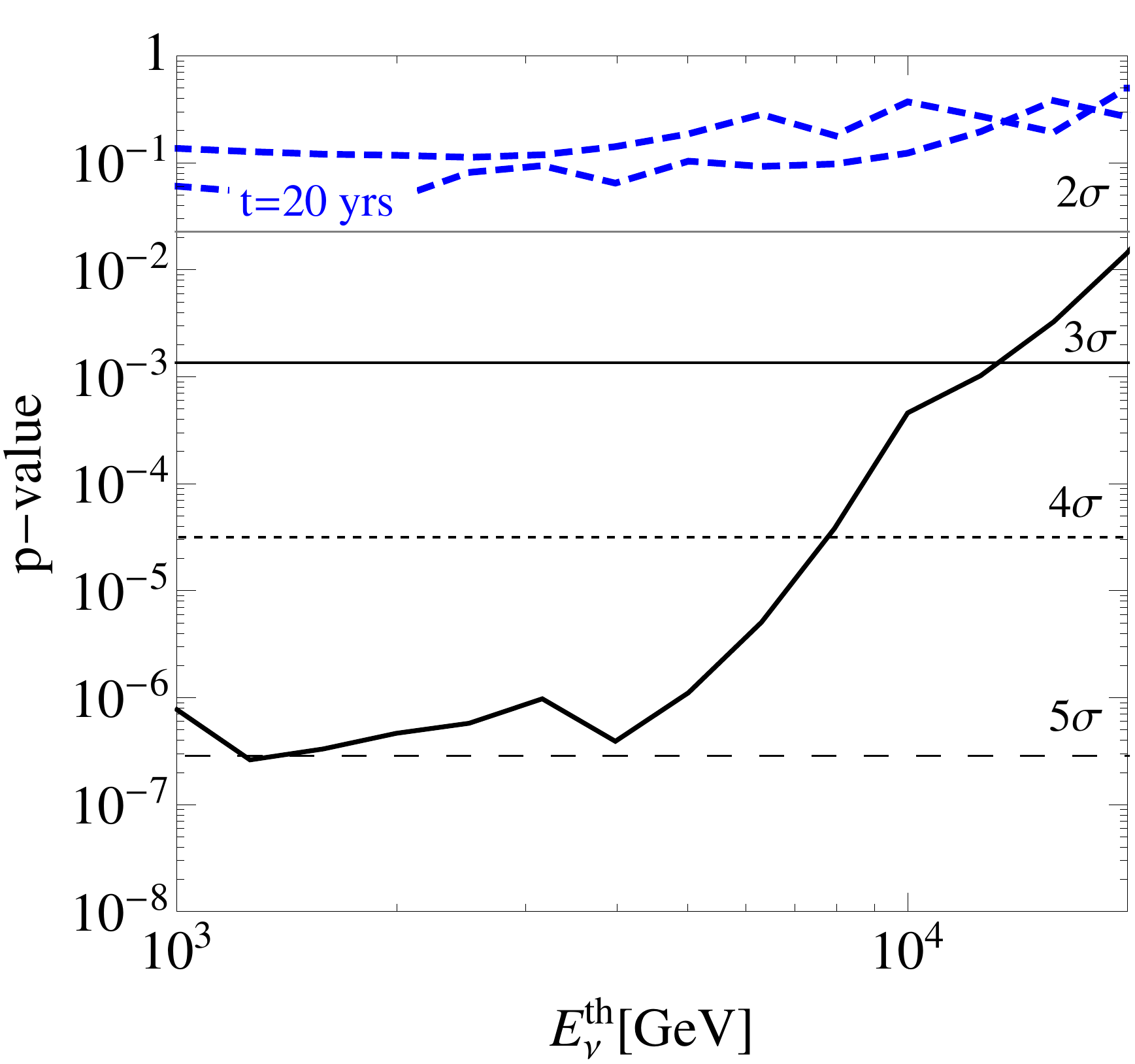}
\caption{p-value as a function of neutrino energy threshold for 10 years running of the 
KM3NeT detector. Different lines have the same meaning as in Fig.~\ref{fig:events}. For the case of 
lepto-hadronic emission, we have considered also 20 years of running time.  }
\label{fig:pvalue}
\end{figure}

In Fig.~\ref{fig:pvalue}, we show the p-value as a function of the neutrino energy threshold. A 5$\sigma$ detection can be reached in 10 years of running of the KM3NeT detector for an energy threshold smaller then 5~TeV, if the gamma-ray emission is fully hadronic. Instead, it is difficult to resolve with a neutrino signal the mixed scenario of leptonic and hadronic emission, considering 10 years of running of the KM3NeT detector. Indeed, for this case the 3$\sigma$ level is not reached in 10~years. However, with a higher running time, of about 20 years, a p-value of the order of several percent could be reached, corresponding to a hint for an excess of neutrinos (a significance approaching 2$\sigma$).
Therefore, in this scenario a detection might be well within the reach of extension beyond the km$^3$ of detectors such as discussed in~\citet{km3net2016}.

\section{Conclusions}

The numerous discussions on the origin of the gamma--ray emission of RX J1713--3946 illustrate the difficulties to interpret which is the content in non--thermal particles accelerated. 
Here we illustrate the importance of  taking into account the reacceleration of pre--existing CR electrons in the case of RX J1713--3946. 
 Remarkably, the amount of reaccelerated particles depends on very little parameters: the density of pre--existing CRs far upstream of the shock, and the minimum momentum above which reacceleration is efficient. 
  However, the amount of particles  accelerated from the thermal pool is typically a fraction of the shock ram pressure $\propto  \rho v_{\rm sh}^2$. Therefore, as a remnant of a core--collapse SN, with the shock wave currently propagating in a low density environment, created during the main sequence of the progenitor star, the situation of RX J1713--3946 provides a case in which the reacceleration of pre--existing CR can become comparable to the fresh acceleration of CRs. Let us also note that the reacceleration of pre--existing CRs does not depend on any CR efficiency. 
 
  Moreover, in this low density environment, the density target material available for proton--proton interaction is reduced, which decreases the amount of gamma rays from hadronic origin, but does not affect the production of gamma rays from leptonic origin. This provides us with a situation where the hadronic gamma--ray signal from freshly accelerated protons is at  a somewhat comparable level to the leptonic gamma--ray signal from leptonic reaccelerated CR electrons. 
 
 This two gamma--ray components naturally produce a broad bump in the GeV to TeV range, with a shape compatible with Fermi--LAT and H.E.S.S. observations. The overall gamma--ray spectrum is obtained with minimal, physically motivated assumptions. In particular 1) the slope of accelerated (and reaccelerated particles) is the one expected in the test--particle case $\propto p^{-4}$, and no additional mechanism producing a deviation from $\propto p^{-4}$ at the shock is needed; 2) pre--existing CR protons and electrons around RX J1713--3946 are described by the local unmodulated CR spectrum derived from measurements at the Earth. 
  
 Finally, we have calculated the number of neutrinos expected from RXJ1713--3946, and estimated the chances of detection with the KM3NeT instrument. In the case of a fully hadronic scenario, the neutrino signal is expected to be a few times above the atmospheric background, making the detection in the $\gtrsim 1$ TeV range possible. In the lepto--hadronic scenario presented above, in which the TeV gamma--ray emission is dominated ICS from reaccelerated electrons, the number of neutrinos expected is reduced and is within the reach of future planned instruments expected to go beyond the km$^3$ detector volume~\citep{Adrian-Martinez:2016fdl}. 

 The maximum momentum of protons is estimated in the RSG wind from the growth of non--resonant streaming instabilities, as given by Eq.~\eqref{eq:pmax}, and when the magnetic field amplification becomes inefficient, using the Hillas criterion. For electrons, the maximum momentum is estimated by equating the acceleration time to the minimum of the synchrotron loss time and the age of the SNR. 
 Alternative hypotheses accounting for a higher level of magnetic field amplification would enhance synchrotron losses, and would make it difficult to explain the gamma--ray spectrum above >10 TeV from ICS of reaccelerated electrons. Moreover, the recipe used to estimate the maximum energy of electrons relies on the hypothesis that the coefficient diffusion is Bohm--like, although other diffusion models are plausible, but would a priori lead to lower maximum momenta~\citep[see e.g., discussion in ][]{cristofari2019}.
 
 The computation of the population of reaccelerated electrons depends on very few ingredients: the fact that the SNR shock is strong, and the presence of CR electrons, considered to be the spectrum measured at the Earth. Let us however mention that if the spectrum of CR protons is known to be somewhat uniform in the Galactic disk, we do not have solid evidence for CR electrons. Therefore, if important variations were present in the CR electron spectrum in different locations of the Galaxy, our calculation might be substantially affected.

\section*{Acknowledgments}
The authors thank P. Blasi for stimulating discussions and comments. 
This project has received funding from the European Union?s Horizon 2020 research and innovation programme under the Marie Sk\l{}odowska-Curie grant agreement No.  843418 (nuHEDGE). 
SG acknowledges support from Agence
Nationale de la Recherche (grant ANR- 17-CE31-0014). SG and PC acknowledge support from the Observatory of Paris (Action Fedératrice CTA). 

\section*{Data Availability}
There are no new data associated with this article.


\bibliographystyle{mnras}
\bibliography{RXJ} 




\appendix

\bsp	
\label{lastpage}
\end{document}